\documentclass[journal,10pt]{IEEEtran}
\usepackage{amsfonts}
\usepackage{amssymb}
\usepackage{stfloats}
\usepackage{cite}
\usepackage{graphicx}
\usepackage{psfrag}
\usepackage{subfigure}
\usepackage{amsmath}
\usepackage{array}
\usepackage{algorithm}
\usepackage{algpseudocode}
\usepackage{setspace}
\usepackage[english]{babel}
\usepackage{blindtext}
\usepackage{url}
\usepackage{epstopdf}
\usepackage{verbatim}
\usepackage{color}
\usepackage{amsmath}
\usepackage{bm}
\usepackage{multirow}
\usepackage{booktabs}
\usepackage{makecell}

\title{First 20 Years of Green Radios}

\author{Shunqing Zhang,~\IEEEmembership{Senior Member, IEEE}, Shugong Xu,~\IEEEmembership{Fellow, IEEE}, \\
Geoffrey Ye Li,~\IEEEmembership{Fellow, IEEE}, and Ender Ayanoglu,~\IEEEmembership{Fellow, IEEE} \\

\thanks{Manuscript received December, 06, 2018; revised March, 31, 2019 and April, 30, 2019; accepted August, 01, 2019.}
\thanks{S. Zhang, S. Xu are with Shanghai Institute for Advanced Communication and Data Science, Shanghai University, Shanghai, 200444, China (e-mail: \{shunqing, shugong\}@shu.edu.cn)}
\thanks{Geoffrey Ye Li is with School of Electrical and Computer Engineering, Georgia Institute of Technology, Atlanta, GA 30332 USA (e-mail: liye@ece.gatech.com)}
\thanks{Ender Ayanoglu is with Department of Electrical Engineering and Computer Science, University of California, Irvine, California, CA 92697 USA (e-mail: ayanoglu@uci.edu)}
\thanks{Corresponding author: Shugong Xu.}
}
\begin{document}
\maketitle

\begin{abstract}
Green radios jointly consider the spectrum efficiency and the energy efficiency in wireless networks in order to provide sustainable development. In the past two decades, various green radio solutions for different types of terminals and radio access networks have been developed, some of which have been used in the designs of wireless products. In this paper, we provide a historical view on some fundamental works and practical issues. In addition to providing a comprehensive overview on theoretical achievements, we also discuss several important power saving solutions with notable engineering impacts, such as Doherty power amplifier and separated baseband unit - remote radio unit architecture, which might be overlooked in previous publications. Moreover, with the huge growth of wireless traffic in the near future, green radio design for future wireless networks shall involve an end-to-end energy efficiency investigation of mobile terminals, radio access and core networks, and different applications. By introducing green radio schemes for advanced terminals and future wireless networks, this article will not only be beneficial for readers with only preliminary background in communications and signal processing but also have reference and historical values for researchers and practical engineers in the area.
\end{abstract}


\section{Introduction} \label{sect:intro}
Information transmission is always associated with energy consumption. According to Shannon's theory, the maximum achievable transmission rate, also known as {\em channel capacity}, scales with the received signal-to-noise ratio (SNR) logarithmically. Traditionally, communication system design focuses on increasing the {\em spectrum efficiency} (SE), i.e., channel capacity per unit bandwidth, with a limited power budget. Towards this goal, various transmission technologies, including orthogonal-frequency-division-multiplexing (OFDM), multiple-input-multiple-output (MIMO), as well as heterogeneous networks, have been proposed to compensate the potential channel fading effects and improve SE.

Green radio jointly considers SE and power consumption to deliver energy efficient wireless communications, where the suggested green metric, e.g., {\em energy efficiency} (EE), is defined as the ratio of the total amount of data delivered and the total energy consumed\footnote{The total amount of data delivered over the total energy consumed is equivalent to the total transmission rate over the total consumed power when the transmission period is fixed. Therefore, spectrum efficiency (normalized total transmission rate) over power is also widely used, such as \cite{Miao10,Chen11}.}. From the definition of EE, green radio can be realized using traditional SE maximization approach under fixed power dissipation \cite{Cui05} or directly minimizing the power consumption while preserving the target transmission rate \cite{Wiart00}. Even if these two approaches do not directly optimize the EE, they were commonly used in the first of the two decades we are considering to design power limited terminals, such as mobile handsets \cite{Perillo04}, sensor nodes \cite{Younis04}, or even relay nodes \cite{Yao05}, and the main efforts focus on maximizing the sleep time through joint design of the hardware \cite{Cui05}, the physical (PHY) layer \cite{Cui04}, the medium access (MAC) layer \cite{Ye04} as well as all other higher layers \cite{Goldsmith02}.

In the second decade after the green radio was proposed, the research focus has moved to the EE study of the radio access network (RAN) side \cite{Miao10,Chen11}, which aims to jointly reduce the power consumption and at the same time improve SE, that is, to maximize the EE of wireless networks. In theory, the fundamental tradeoff relation between SE and EE has been discovered in \cite{Miao10} to show that SE and EE cannot be improved {\em simultaneously} if the practical circuit power consumption is considered. Similar conclusions can be applied to deployment efficiency (DE) versus EE, delay versus power, and bandwidth versus power as shown in \cite{Chen11}. Guided by this green radio tradeoff framework, the energy efficient schemes for RANs concentrate on exploiting different network operation configurations \cite{Xu13} rather than directly turning into sleep mode, as frequently adopted in power limited terminals. In addition, some breakthroughs in engineering, e.g., Doherty architecture for power amplifiers (PA) \cite{Bumman06}, and separated baseband unit (BBU) and radio remote unit (RRU) architecture \cite{Wang09}, greatly contribute to the development of the green access networks.

Nowadays, due to the huge growth of wireless traffic and the corresponding energy consumption explosion, operating wireless communication networks without green radio technologies is no longer sustainable, which triggers EE to be selected as one of the basic system performance measures for the Fifth Generation (5G) wireless communication networks \cite{I14}. Different from the past 20 years, green radio design for 5G wireless networks involves end-to-end energy efficient mobile terminals, access, core networks, and different applications. Some design challenges are listed as follows.
\begin{itemize}
\item First, the major application scenarios defined by 5G networks \cite{5G17} include enhanced mobile broadband (eMBB), ultra-reliable and low-latency communications (URLLC), and massive machine type communication (mMTC), which requires a new definition of green radio to handle massive connectivity, latency, and reliability.
\item Second, since 5G networks extend the traditional people-to-people communication to people-to-machine and machine-to-machine communications, new types of mobile terminals, e.g. vehicles \cite{Liang17,Peng18,Qin18} or renewable energy empowered devices \cite{Jiang16}, shall be supported and the corresponding energy efficient solutions need to be explored.
\item Last but not least, with massively deployed network nodes, such as massive MIMO \cite{Lu14} and ultra dense networks (UDN) \cite{Yunas15}, the information exchange among massive network nodes is no longer negligible and incurs new design challenges for energy efficient schemes. It can be expected that with more than 50 billion mobile terminals supported and billions of network nodes deployed in the near future, green radio will play a more critical role in wireless system design and deployment of the future.
\end{itemize}

In this paper, we provide a historical view on some fundamental issues and important practical issues that contribute to the current green radio design of long term evolution (LTE) and 5G communication systems rather than just providing an overview of the theoretical progress and specific technologies in green radio \cite{Zhang17,Wu17,Wu15}. Specifically, in addition to providing a comprehensive study of theoretical achievements, such as fundamental tradeoffs \cite{Miao10,Chen11} or energy efficient scheduling schemes \cite{Xiao06,Xiong11}, we also discuss several power saving solutions with notable engineering impacts, including but not limited to discontinuous transmission \cite{Perillo04} and reception \cite{Yang05}, Doherty PA \cite{Bumman06}, and separated BBU-RRU architecture \cite{Wang09}, which might be overlooked in many past survey or tutorial papers. By jointly elaborating the progresses of the green radio in its first 20 years and forecasting some open issues, this article will not only be beneficial for the readers with only preliminary background in communications and signal processing, but also have reference and historical values for researchers and practical engineers in the area.

The rest of the paper is organized as follows. In Section~\ref{sect:gr}, we introduce the fundamental theory and briefly discuss the milestones in the history of green radios. In Section~\ref{sect:tt}, we present green radio solutions for battery-powered terminals to prolong standby time. In Section~\ref{sect:an}, we focus on the green radio schemes for access networks to reduce the energy consumption and maintain a sustainable operation cost. Section~\ref{sect:at} and Section~\ref{sect:fn} provide a comprehensive overview of green radio technologies for future massive wireless terminals and networks, respectively. Final remarks are given in Section~\ref{sect:conc}.

A list of abbreviations and acronyms used in this paper are provided in Table~\ref{tab:abrv}.
\begin{table*} [ht]
\centering
\caption{Summary of Abbreviations and Acronyms Used in This Paper.}
\label{tab:abrv}
\footnotesize
\begin{tabular}{c c c c}
\toprule
Abbr./Acron. & Definition & Abbr./Acron. & Definition\\
\midrule
3GPP & the third generation partnership project  & MEC & mobile edge computing \\
\midrule
5G & the fifth generation & MBSFN & multicast broadcast single frequency network \\
\midrule
ABS & almost blank subframe & MIMO & multiple-input-multiple-output \\
\midrule
ADC & analog-to-digital converter & mMTC & massive machine type communications\\
\midrule
BB & baseband & mmWave & millimeter wave\\
\midrule
BBU & baseband unit & NB-IoT & narrow band IoT \\
\midrule
BS & base station & NET & network \\
\midrule
CoMP & coordinated multi-point & NFV & network function virtualization \\
\midrule
CPRI & common public radio interface & OFDM & orthogonal-frequency-division-multiplexing \\
\midrule
C-RAN & cloud RAN & PA & power amplifier\\
\midrule
D2D & device-to-device & PAPR & peak-to-average power ratio \\
\midrule
DE & deployment efficiency & PHY & physical \\
\midrule
DPD & digital pre-distortion & QoS & quality of services \\
\midrule
DRX & discontinuous reception & RAN & radio access network \\
\midrule
DTX & discontinuous transmission & RF & radio frequency \\
\midrule
EE & energy efficiency & RFID & radio frequency identification \\
\midrule
eICIC & enhanced inter-cell interference coordination & RRU & remote radio unit \\
\midrule
eMBB & enhanced mobile broadband & RS & reference signal \\
\midrule
GSM & global system for mobile communications & SDN & software defined network \\
\midrule
IIoT & Industrial IoT & SE & spectrum efficiency \\
\midrule
IoT & Internet of Things & SNR & signal-to-noise ratio \\
\midrule
LTE  & long time evolution & UDN & ultra dense networks \\
\midrule
LTE-U & LTE unlicensed & UMTS & universal mobile telecommunications systems \\
\midrule
LPWA & low power wide area & URLLC & ultra-reliable and low-latency communications \\
\midrule
MBSFN & multicast broadcast single frequency network & V2V & vehicle-to-vehicle \\
\bottomrule
\end{tabular}
\end{table*}

\section{Overview of Green Radios} \label{sect:gr}

\begin{figure}
\centering
\includegraphics[width = 3 in]{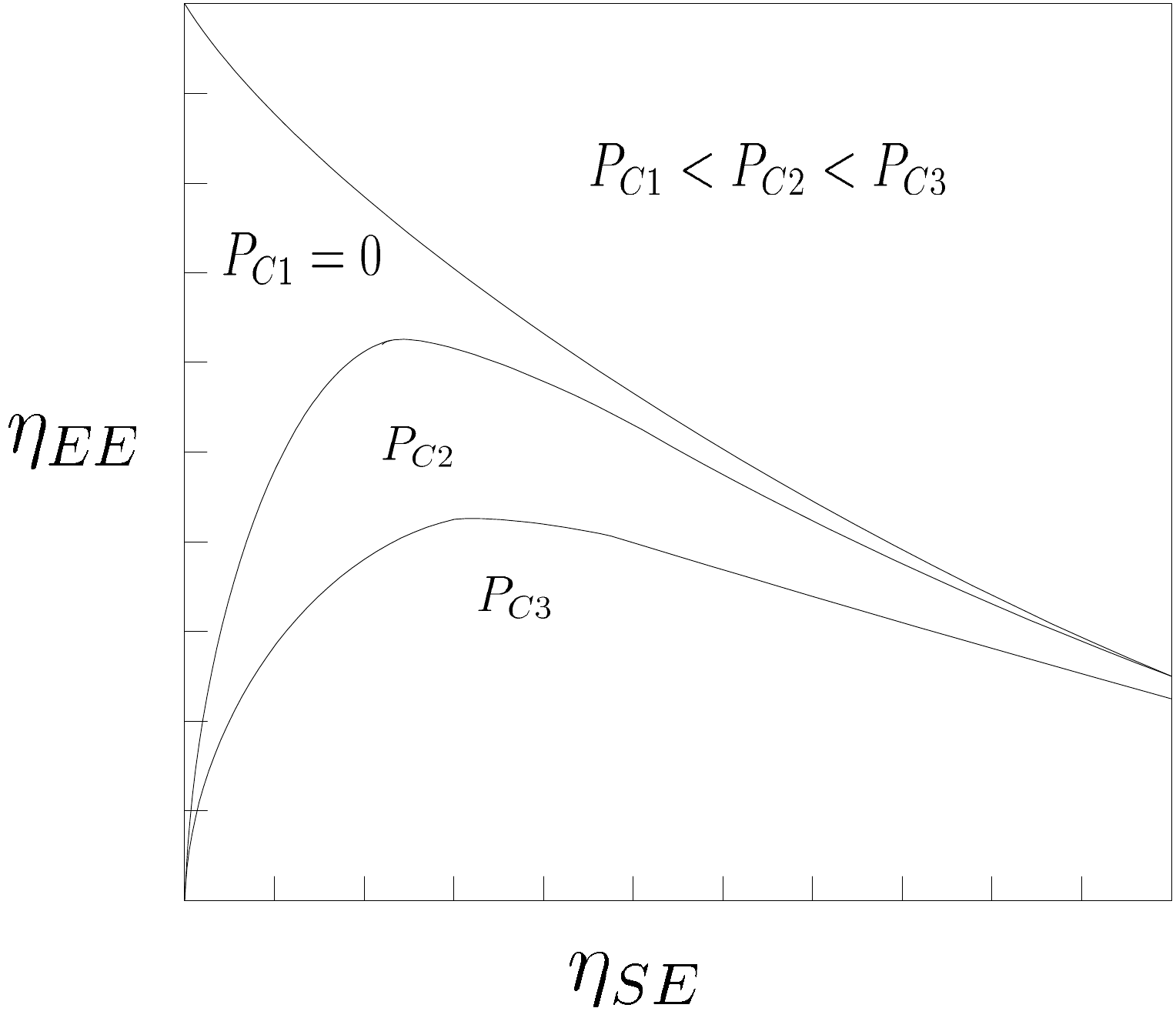}
\caption{The SE-EE tradeoff curves under different circuit powers \cite{Li11}.}
\label{fig:SE-EE}
\end{figure}

Green radios involve lots of parameters and metrics in wireless networks, widely spread in the network operation and maintenance. Although there are many tradeoffs among them, four fundamental tradeoffs have been identified in \cite{Chen11}, which systematically characterize the most critical relations in the green aspect. The SE-EE tradeoff is the most popular one among them, which balances the system achievable throughput and the corresponding energy consumption. The mathematical expression for the SE-EE tradeoff under the additive white Gaussian noise environment is given by \cite{Zhang17},
\begin{eqnarray} \label{eqn:EE}
\eta_{EE}(\eta_{SE}) = \frac{\eta_{SE}}{(2^{\eta_{SE}} - 1) N_0 + P_{c}},
\end{eqnarray}
where $N_0$ denotes the additive noise power density, $P_{c}$ denotes the circuit power consumption, and $\eta_{SE}, \eta_{EE}$ denote the system SE and EE, respectively. Figure~\ref{fig:SE-EE} shows $\eta_{EE}$ versus $\eta_{SE}$ for different circuit powers. From equation \eqref{eqn:EE} and Figure~\ref{fig:SE-EE}, we can draw the following conclusions.
\begin{itemize}
\item In the ideal case, $P_{c} = 0$, without the consideration of the circuit power, EE decreases monotonically with SE and they cannot be optimized simultaneously.
\item In the practical case, $P_{c} > 0$, considering the circuit power, the SE-EE tradeoff behaves like a bell shape \cite{Li11} and the optimal point to operate the wireless network with the maximum EE exists. Meanwhile, the SE-EE tradeoff region will be enlarged with the reduced circuit power (e.g. smaller $P_{c}$).
\end{itemize}

Based on the above understanding, green radios to achieve better EE for the given throughput requirement can be classified into the following two categories, where the system SE, $\eta_{SE}$, the circuit power, $P_{c}$, and other important factors to affect EE are jointly considered.
\begin{itemize}
\item{\em Moving Towards the Maximum EE Point.} The most common approach to move towards the maximum EE point is through energy efficient scheduling \cite{Xiao06}. Rather than to operate the system at the maximum SE point as in the traditional case, the energy efficient scheduling schemes adjust spatial \cite{Xu13}, time and frequency \cite{TChen11}, as well as power resources \cite{Xiong11} to achieve a better EE result.
\item{\em Enlarging the SE-EE Tradeoff Region.} One straightforward solution to enlarge the SE-EE tradeoff region is to reduce the circuit power, $P_{c}$, as mentioned before, which covers a wide range of engineering efforts, including the discontinuous operating of circuits \cite{Perillo04,Yang05} and the highly efficient PA architecture \cite{Bumman06}. Another way is via cooperations. Typical solutions include using UDN \cite{Yunas15} and device-to-device (D2D) communications \cite{Yu11,Feng13,Feng14} to shorten the equivalent communication distances and applying collaborative transmission \cite{Cui04} to improve the equivalent SNR.
\end{itemize}

The above methodology can be applied to other tradeoffs as well, such as the DE-EE tradeoff. For example, the energy harvesting technologies have been recognized as sustainable solutions for wireless networks by exploring new types of energy, such as renewable resources or RF signals \cite{Wu17}. Although the overall SE-EE tradeoff is not affected, the corresponding energy costs will be reduced and the DE-EE tradeoff can be enlarged. This is because part of the total energy can be supported by renewable sources.

There has been tremendous achievement in green radios in the past 20 years. In Table~\ref{tab:milestone}, we summarize the milestones of green radios during the first 20 years. In addition to that, part of research efforts have also been contributed to the backbone infrastructure, also known as core networks of wireless communications, including network routers, optical cross-connect, and transponders. Typical green solutions for them include the energy efficient core network architecture design \cite{Shen09}, efficient routing and packet forwarding protocol design \cite{Buysse13}, and selectively turning off idle elements \cite{Zhang11}. Since the major energy consumption of ``radios'' is in the access network and terminal sides, we refer interested readers to reference \cite{Zhang10} for more information on the energy efficient design for core networks.

\begin{table*} [ht]
\centering
\caption{List of Milestones for Green Radios during the First 20 Years.}
\label{tab:milestone}
\footnotesize
\begin{tabular}{c c c}
\toprule
Year & Contributions & Reference \\
\midrule
2000 & Discontinuous transmission for GSM has been proposed and analyzed. & \cite{Wiart00} \\
\midrule
2002 & EE has been defined. & \cite{Goldsmith02} \\
\midrule
2004 & EE of MIMO and cooperative MIMO has been analyzed. & \cite{Cui04}\\
\midrule
2005 & Energy efficient link adaptations have been proposed. & \cite{Cui05} \\
\midrule
2010 & The SE-EE tradeoff has been discovered. & \cite{Miao10}\\
\midrule
2011 & Four fundamental tradeoffs for green radios have been discovered. & \cite{Chen11}\\
\midrule
2013 & D2D communications have been used for green radios. & \cite{Feng13} \\
\midrule
2014 & Energy efficient massive MIMO schemes have been proposed. & \cite{Bjornson14} \\
\midrule
2015 & Energy harvesting for IoT has been proposed. & \cite{Kamalinejad15} \\
\midrule
2017 & Energy efficient vehicular communications have been proposed. & \cite{Liang17} \\
\bottomrule
\end{tabular}
\end{table*}

\section{Traditional Terminals} \label{sect:tt}

As we have mentioned before, the main focus of green radios before 2008 is on the terminal side since the battery limitation is a critical consideration for designing traditional terminals, such as mobile handsets and sensor nodes, at that time.

Traditional terminals consist of most blocks including duplexers, mixers, filters, and other auxiliary components, each with certain power consumption. Since most of the above components are tightly coupled with each other, the main approach to save energy is to turn off the whole transmitting or receiving chains and enter the terminal to the {\em sleep mode} for as long as possible. Depending on the types of terminals, green radios can be mainly categorized into two classes, namely {\em centralized-controlled} and {\em self-organized}. Since the centralized-controlled design needs to follow strict regulations by the standards, self-organized approaches have more flexibility in EE improvement.

\subsection{Mobile Handsets}
Green radios for mobile handsets can be classified as the centralized-controlled solution, where every mobile handset needs to exactly follow the procedures defined by the standard, such as global system for mobile communications (GSM) or the third generation partnership project (3GPP) universal mobile telecommunications systems (UMTS). According to the standard specifications, base stations (BSs) shall broadcast important system information periodically and the terminals are required to receive this message and update their local databases correspondingly. Therefore, green radio approaches for mobile handsets have to sleep and wake up periodically, which is known as {\em discontinuous reception} (DRX), and the corresponding period is often called the {\em DRX cycle}.

\begin{figure*}
\centering
\includegraphics[width = 6 in]{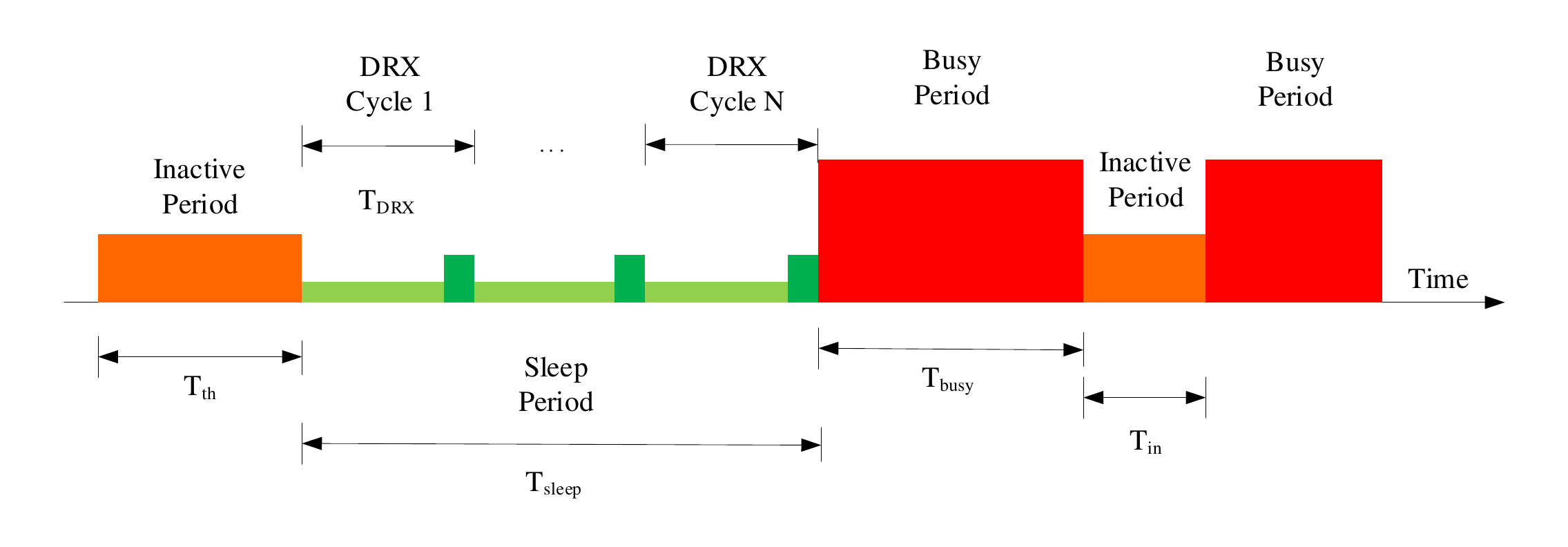}
\caption{An illustrative example of discontinuous reception (DRX) scheme for mobile handsets. In this example, the threshold for the inactivity timer is $T_{th}$, the duration of DRX cycle is $T_{DRX}$, and the durations of busy, inactive, and sleep periods are labeled with $T_{busy}$, $T_{in}$, and $T_{sleep}$, respectively.}
\label{fig:tt_drx}
\end{figure*}

As shown in Figure~\ref{fig:tt_drx}, the mobile handsets will listen to the wireless medium in the {\em inactivity} period and start the timer after the existing packet transmission is finished. If a new packet arrives before the timer expires, mobile handsets will turn into the {\em busy} period for packet receiving and reset the timer afterwards. Otherwise, mobile handsets will enter the {\em sleep} period and the receiving chains will be switched off. According to the 3GPP specifications \cite{DRX12}, the sleep period contains at least one DRX cycle and the mobile handsets will wake up to check the status by listening to the paging information\footnote{The paging information is usually contained in the paging channel for GSM and UMTS systems.} at the end of each DRX cycle. The sleep period will continue if there is no package arrival.

DRX scheme is quite efficient from the engineering point of view. The corresponding analysis on the power saving gain and the potential effects on quality of services (QoS) can be found in \cite{Yang05} and \cite{Xiao05}. The analytical relations among the design parameters (e.g., the inactivity timer threshold and the duration of the DRX cycle), the effects on QoS (e.g., the expected queue length and the expected packet waiting time), and the power saving factor have been provided in \cite{Yang05}. A similar framework can be also applied to IEEE 802.16e systems as well for quantitative energy saving analysis \cite{Xiao05}.

\subsection{Sensor Nodes}
Wireless sensor networks are equipped with a self-organized protocol and the system information update for sensor nodes is basically event-driven rather than periodic. Therefore, sensor nodes apply a different design philosophy from other wireless terminals and deliver sensing reports more efficiently by optimizing the transmission strategies in different layers, in order to prolong the potential sleeping period.

In the PHY layer, the straightforward way to deliver green radio is to design efficient transmission schemes, such as modulation and coding, under certain energy constraints. As shown in \cite{Cui05}, if we consider both the transmission energy and the circuit energy consumption, nearly 80\% of the total energy consumption can be saved by utilizing adaptive modulation and coding schemes. Collaborative MIMO transmission has been proposed in \cite{Cui04} as another possible PHY layer solution to improve EE. By sharing sensing reports among different sensors, a simple collaborative multi-sensor transmission strategy using the {\em Alamouti} scheme \cite{Alamouti98} can easily beat traditional single sensor approach when the transmission distance is greater than a threshold. If adaptive modulation schemes can be deployed, the threshold can be further reduced from $62m/75m$ to $3.8m/1.6m$ for antenna configurations $2\times1$ and $2\times2$, respectively \cite{Cui04}.

In the MAC layer, the fundamental task is to avoid potential collisions in the multi-sensor environment. Since periodic system information broadcasting used in a centralized manner is usually unavailable in sensor networks and time division or frequency division based multiple access schemes are not supported in general, a contention based protocol, such as in IEEE 802.11, must be used and incurs significant energy waste due to the unorganized behaviors including carrier sensing, overhearing, and collision. To address this issue, coordinated adaptive sleeping \cite{Ye04} has been proposed, where sensor nodes are required to enter the sleep mode periodically and the duty cycle (similar to the DRX cycle concept for mobile handsets) information is shared among the neighboring sensors. Through this approach, a group of nearby sensors shall sleep according to the same duty cycle in order to mitigate the potential collisions. In addition, if the neighboring sensors need to communicate with each other, wake-up duration in the duty cycle can be utilized, which prevents additional energy consumption in the contention process.

In the network (NET) layer, in order to reduce the power consumption of sensor networks, efficient routing protocols, such as data-centric routing and clustering-based routing, are commonly used. Although the power consumption overhead associated with data-centric routing discovery and cluster information processing is usually significant, a dynamic clustering protocol with centralized control and randomized rotation of cluster head proposed in \cite{Muruganathan05} provides energy efficient information exchange among different sensor nodes. Without centralized control, energy efficient routing schemes can also be implemented in a distributed manner. By jointly considering direct transmission and cooperative transmission, the resultant distributed cooperative routing scheme in \cite{Ibrahim08} is able to achieve more than 50\% power saving compared with the traditional shortest path routing algorithms.

Besides the above solutions, the green radio schemes can exploit specific features in some typical application scenarios. For example, in \cite{Liu07}, the spatial and temporal correlations among sensing results are exploited to save energy. In \cite{Lin09}, energy efficient distributed adaptive sensing utilizes the continuous trajectory property in the target tracking case. In Table~\ref{tab:term}, we summarize the green radio schemes for traditional terminals.

\begin{table*} [ht]
\centering
\caption{Summary of Green Radios for Traditional Terminals.}
\label{tab:term}
\footnotesize
\begin{tabular}{c c c}
\toprule
Green Radio Schemes & Description & Reference \\
\midrule
DRX & periodically enter the sleep mode when DRX is activated & \cite{DRX12} \\
\midrule
Efficient Transmission & energy constrained adaptive modulation, coding, and collaborative transmission & \cite{Cui05,Cui04} \\
\midrule
Collision Avoidance & coordinated adaptive sleeping and transmission & \cite{Ye04} \\
\midrule
Efficient Routing & dynamic clustering based routing and distributed cooperative routing & \cite{Muruganathan05,Ibrahim08} \\
\midrule
Feature Exploiting & exploit the correlated sensing results and the continuous trajectory property & \cite{Liu07,Lin09} \\
\bottomrule
\end{tabular}
\end{table*}

\section{Radio Access Networks} \label{sect:an}
\begin{figure}
\centering
\includegraphics[width = 3.4 in]{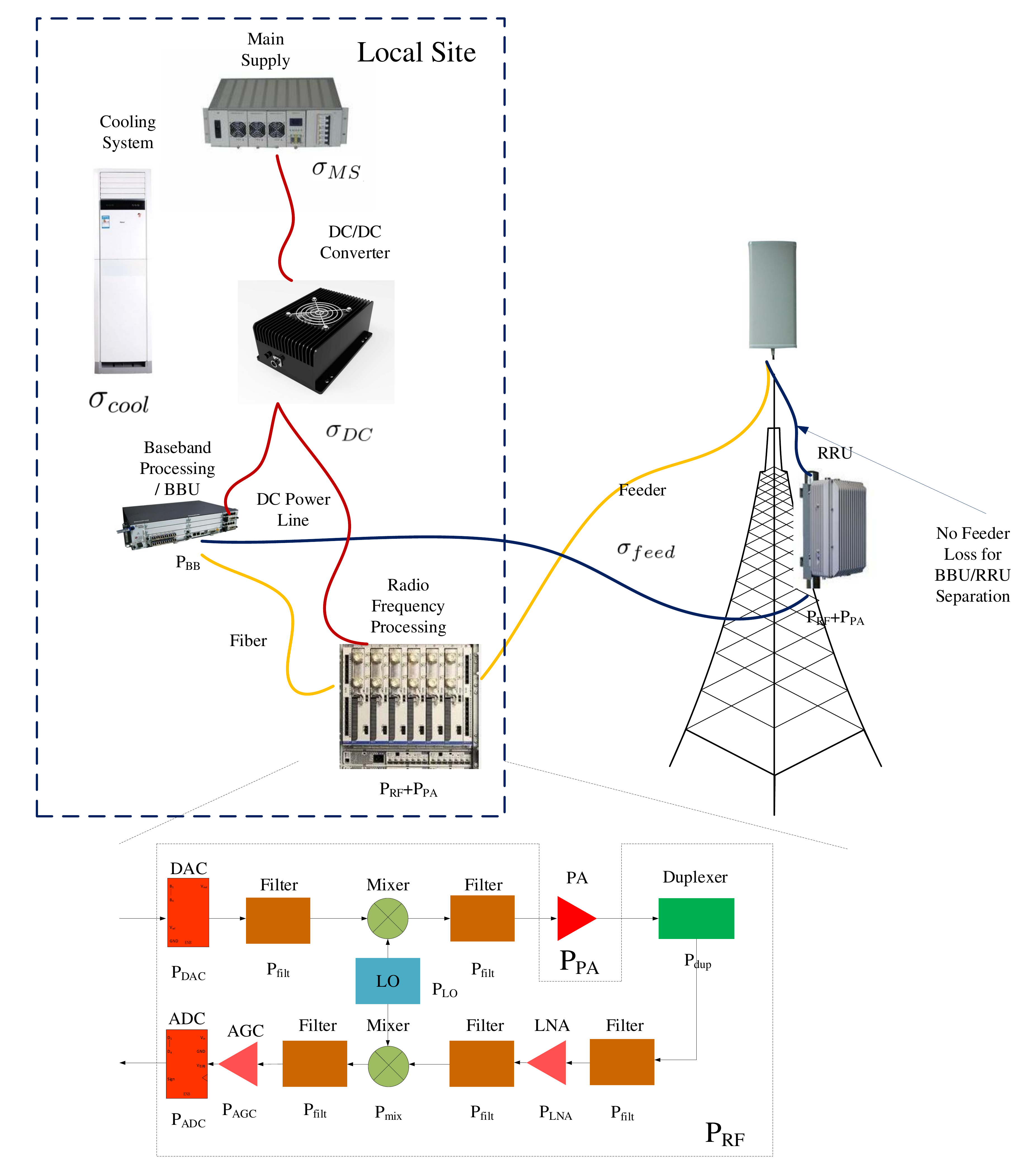}
\caption{The abstracted power model for radio access networks, including the coupling losses from the feeders, the DC-DC converter, the main supply, and the cooling systems. Different from the power model of traditional terminals, a constant value of $P_{RF}$ is assumed for RF processing (except for PA), since the power consumption variations of RF components are relatively small, if compared with PA part.}
\label{fig:net_model}
\end{figure}

Due to the divergent trends between the network operation revenue and the operational expenditure, green radios have switched the focus from terminal side to RAN side after 2008. The corresponding power consumption at the BSs\footnote{Since the number of BSs is much larger than other equipments in the radio access networks, the green radios for radio access networks mainly focus on the energy saving schemes for BSs. }, as shown in Figure~\ref{fig:net_model}, is proportional to the number of transceiver chains, $N_{\textrm{tx}}$, and the total power consumption, $P_{t}$, can be expressed as \cite{EARTH:D2.3},
\begin{eqnarray} \label{eqn:bs_mod}
P_{t} = N_{\textrm{tx}}\cdot \frac{\frac{P_{o}}{\eta_{PA} \left(1 - \sigma_{feed}\right)} + P_{RF} + P_{BB}}{\left(1 - \sigma_{DC}\right)\left(1 - \sigma_{MS}\right)\left(1 - \sigma_{cool}\right)},
\end{eqnarray}
where $\sigma_{feed}$, $\sigma_{DC}$, $\sigma_{MS}$, and $\sigma_{cool}$ denote the coupling losses from the feeders, the DC-DC converter, the main supply, and the cooling systems, respectively, $P_{o}$ represents the average transmit power for each transceiver link, $P_{RF}$ and $R_{BB}$ are the power consumptions of the baseband (BB) and radio frequency (RF) processing, respectively. To simplify green radio analysis for RANs, the following linear approximation is often applied to the above model,
\begin{eqnarray} \label{eqn:bs_mod_simp}
P_{t} = \left\{
\begin{array}{l l}
N_{\textrm{tx}} \cdot \left(P_{\textrm{sta}} + \Delta_{p} P_{0} \right), & 0 < P_{o} \leq P_{\max},\\
N_{\textrm{tx}} \cdot P_{\textrm{sleep}}, & P_{o} = 0,
\end{array}
\right.
\end{eqnarray}
where $P_{\textrm{sta}}$ and $P_{\textrm{sleep}}$ denote the power consumption in the idle and sleep mode respectively. $\Delta_{p}$ is the associated linear coefficient and $P_{\max}$ is the maximum transmit power per each link. Different from the traditional terminals, the dynamic range of power consumptions for RANs is no longer negligible. With different types of BSs in the practical deployment, the corresponding parameters are summarized in Table~\ref{tab:para}.

\begin{table} [ht]
\centering
\caption{Power Model Parameters for Different BS Types \cite{EARTH:D2.3}.}
\label{tab:para}
\footnotesize
\begin{tabular}{c c c c c c}
\toprule
\textbf{BS Type}&$N_{\textrm{tx}}$&$P_{\max}$ (w) & $P_{\textrm{sta}}$ (w) & $\Delta_{p}$ & $P_{\textrm{sleep}}$ (w)\\
\midrule
Macro Cell & 6 & 20 & 130.0 & 4.7 & 75.0\\
\midrule
Radio Remote Head & 6 & 20 & 84.0 & 2.8 & 56.0\\
\midrule
Micro Cell & 2 & 6.3 & 56.0 & 2.6 & 39.0\\
\midrule
Pico Cell & 2 & 0.13 & 6.8 & 4.0 & 4.3\\
\midrule
Femto Cell & 2 & 0.05 & 4.8 & 8.0 & 2.9\\
\bottomrule
\end{tabular}
\end{table}

Green radios for RANs, therefore, focus on directly minimizing the parameters in the power model in \eqref{eqn:bs_mod} through spatial, time, and frequency domain strategies. Besides the above techniques, several engineering considerations for power saving will be also discussed in this section, including improving the PA efficiency and reducing the coupling losses, such as the feeder loss.

\subsection{Spatial Domain Strategy}

One of the most common approaches to realize the spatial domain strategy for green RAN is to reduce the equivalent number of transceiver chains, $N_{\textrm{tx}}$, by configuring part of them into non-active status. For example, the energy efficient resource allocation scheme for the general MIMO systems in \cite{Xu13} dynamically turns off transceiver chains in the low traffic region, which can be optimized by scaling the number of the active RF chains with the required SE. Since the supported number of spatial streams in the wireless fading environment is in general limited, especially in the massive MIMO scenarios, a better solution is to adopt the hybrid precoding structure with a limited number of RF processing chains \cite{Han15,Gao16}. As proposed in \cite{Liu14}, with a high-dimensional phase-only RF precoder and a low-dimensional BB precoder, the minimum average data rate of users as well as the minimum per-user EE can be maximized. Furthermore, with sufficiently large degree of freedom provided by massive MIMO systems, the hardware impairments with a limited number of RF chains will no longer be an issue any more, which allows to use inexpensive and energy efficient antenna components in the practical deployment \cite{Bjornson14}. It is worth to note that the cell coverage is not an issue any more with MIMO techniques since it is guaranteed by the primary antenna with the maximum transmit power in the practical deployment.

Heterogeneous networks aim to boost capacity by reducing the network size, which have been identified as the simplest and most effective way for EE improvement \cite{Hoydis11}. Although similar ideas to turn off some active cells in the low traffic region can be utilized for EE improvement, the most challenging issue is how to guarantee the full network coverage after cell switch-off. Initially, small cells are within the coverage of always-on macro cells for the hierarchical cell structure \cite{TChen11}, and an additional procedure is to migrate the associated user terminals to macro cells through cell reselection before turning off small cells. Afterwards, coordinate cell zooming technique has been proposed in \cite{Niu10} to compensate the potential coverage holes due to cell sleeping, where the transmit power and antenna tilts from neighboring cells can be adjusted and coordinate multi-point (CoMP) transmission can be applied. A more comprehensive study of the cell switch-off is provided in \cite{Oh13}, where the locations and the load balancing effects of BSs are jointly considered. By taking the amount of offloaded traffic for neighbouring cells into account, the proposed solution turns off BSs one by one in order to minimize the negative effects of wireless networks and achieves 50-80\% power savings under a real traffic profile. In \cite{Soh13}, the corresponding analytical framework for network EE is provided. From \cite{Soh13}, heterogeneous networks generally lead to higher EE and the improvement saturates only if the density of small cells tends to infinity.

\subsection{Time Domain Strategy}

\begin{figure*}
\centering
\includegraphics[trim = 0mm 150mm 0mm 20mm, clip=true, width = 6 in]{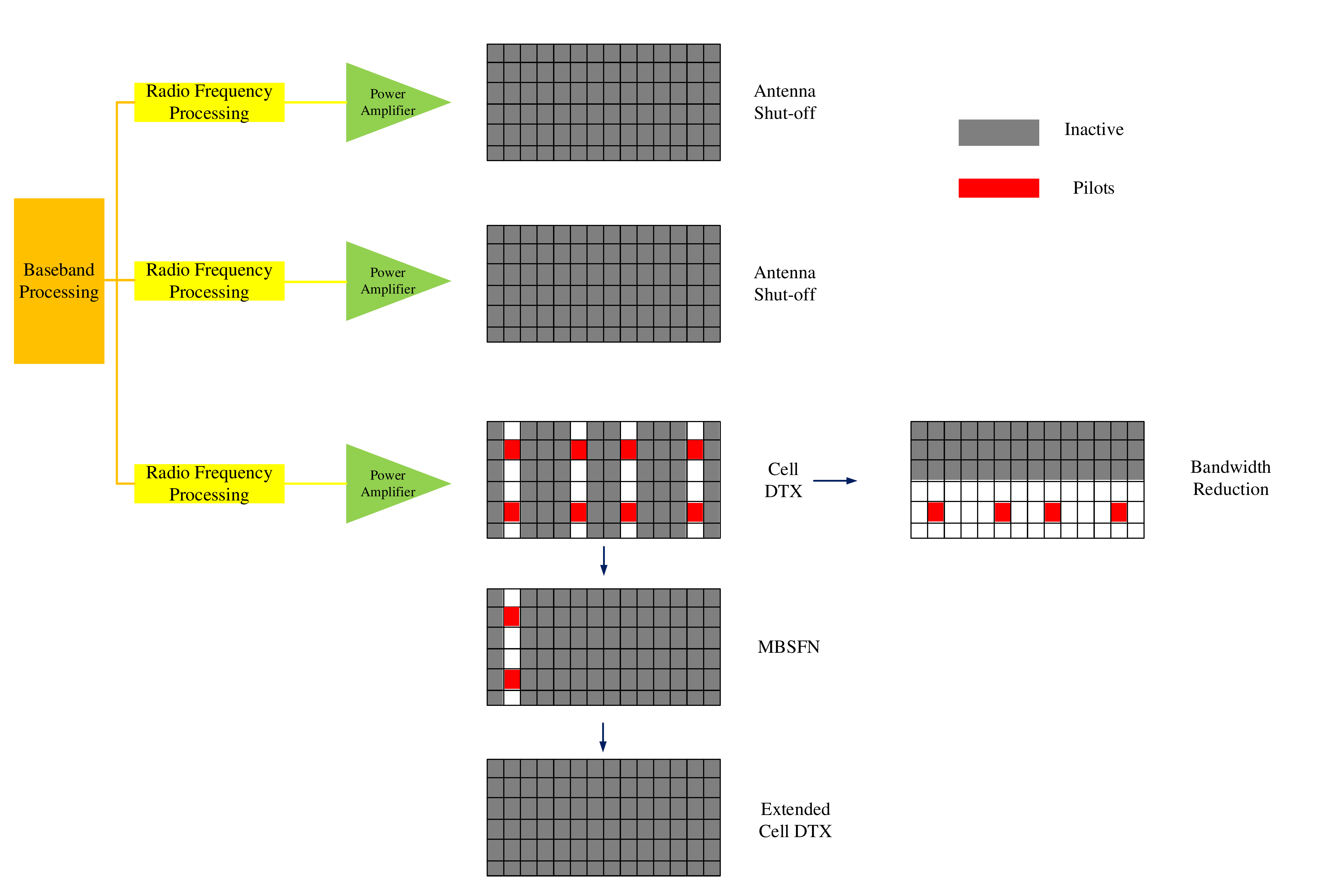}
\caption{An overview of green radios for RANs, including spatial domain (antenna shut-off), time domain (cell DTX), and frequency domain (bandwidth reduction) strategies.}
\label{fig:BS_es}
\end{figure*}

In addition to minimizing the number of active transceiver chains, another possible solution to control the BS power consumption is to extend the durations of sleep modes as indicated by the simplified power model in \eqref{eqn:bs_mod_simp}. Since the slot level PA shut-off is commonly deployed in modern wireless communication networks, the cell discontinuous transmission (DTX) technique to timely switch off the traffic channels as well as the related PAs on a slot basis can significantly prolong the sleeping periods, and therefore, is considered to be an efficient approach for green radios \cite{TChen11}. Conventional cell DTX is realized by turning off PAs at signal-free slots as shown in Figure~\ref{fig:BS_es}. For example, in a normal LTE subframe with 14 OFDM symbols, only four of them are used for reference signals (RSs) and the remaining symbols can be signal-free when the cell is idle. A more aggressive strategy is to configure the normal LTE subframe into multicast broadcast single frequency network (MBSFN) subframe, where only one OFDM symbol is needed for RS transmission in each subframe as shown in Figure~\ref{fig:BS_es}. Through this backward compatible solution, an additional three OFDM symbols can be utilized for the PA shut-off. To further reduce the remaining RSs, the extended cell DTX concept with no RS transmission has been proposed in \cite{DTX09} as an energy saving feature. Since this idea requires additional control signaling support from the higher layers, it is not accepted by the 3GPP standard until the ``almost blank subframe (ABS)'' configuration becomes part of enhanced inter-cell interference coordination (eICIC) feature in 3GPP LTE Release 10 \cite{ABS12}.

To efficiently utilize the potential power saving gains from cell DTX or extended cell DTX, a joint consideration with power control and spatial domain adaptation is often required. In \cite{Holtkamp14}, the proposed joint algorithm is able to reduce the overall power consumption by a factor of 25\% to 40\% in a single cell case compared with the separated schemes. For multi-cell scenario, a joint macro cell muting using extended cell DTX to improve user experiences according to user mobility and traffic dynamics has been studied in \cite{Vasudevan13}.

\subsection{Frequency Domain Strategy}
Another straightforward approach to minimize the power consumption of BS is to reduce the transmit power requirement, $P_{o}$, through the bandwidth reduction technique in \cite{TChen11}. This type of energy saving is due to the following two reasons. First, the radiated power generally scales with the bandwidth requirement since the power spectral density is maintained in the modern wireless systems. Second, smaller bandwidth requires less reference signals, which further reduces the transmit power budget. Intuitively, we can achieve the maximum power saving gain through bandwidth reduction if the required bandwidth can scale with the traffic load. However, it requires control signaling support from the higher layers\footnote{For example, a system bandwidth update message needs to be broadcasted to all the terminals within the cell, when the bandwidth reduction happens.} and cannot be applied very frequently. Meanwhile, the power saving gain is often compromised as the transmit power reduction may result in a degradation of the normal PA efficiency.

To fully exploit the frequency domain power saving gain, throughput and power requirements have been jointly considered in \cite{Miao10} in order to maximize the overall EE in frequency selective channels. By allocating the amount of power according to the throughput requirement as well as the responses of subchannels in OFDM systems, the energy efficient link adaptation algorithm achieves more than 15\% EE improvement. A similar phenomenon has been observed in the multiuser environment as well \cite{Xiong11,Xiong12}, where a more sophisticated EE-SE tradeoff is evaluated under the priority and fairness constraints.

In addition to making full utilization of available cellular bands, another way is to explore additional spectrum, such as millimeter wave (mmWave) or unlicensed bands. As the expansion of extra bands is able to provide a linear capacity boost with respect to the available bandwidth, the related energy efficient schemes concentrate on providing low cost and low power implementations. For example, as the wavelength of a mmWave carrier is much smaller, a large number of antenna elements are often equipped to achieve large array gains \cite{Han15,Gao16}. To reduce the implementation cost, 5GHz WiFi band with LTE air interface is selected for LTE unlicensed (LTE-U) application according to 3GPP Release 13 \cite{Wu17}.

\subsection{Engineering Aspects}
Apart from the systematical power saving schemes, the engineering consideration of implementing BSs is also critical in green radios. Based on the BS power model \eqref{eqn:bs_mod}, the PA efficiency, $\eta_{PA}$, as well as the coupling losses, including $\sigma_{feed}$ and $\sigma_{cool}$\footnote{The coupling losses of DC-DC converter ($\sigma_{DC}$) and main supply are less than 10\% according to the current technology, which is much smaller than $\sigma_{feed}$ and $\sigma_{cool}$. Therefore, the current green radio design focuses on reducing the values of $\sigma_{feed}$ and $\sigma_{cool}$.}, can be improved with engineering evolution. To improve the PA efficiency, several advanced PA architectures have been proposed. For example, the envelop-tracking architecture has been developed in \cite{Wang05} for OFDM systems by envelope elimination and restoration according to the output power profiles, which achieves 30\% PA efficiency at 2.4 GHz band. To enlarge the maximum output power of PAs, Doherty architecture, consisting a carrier amplifier for low to medium output power region and a peak amplifier for high output power region, has been widely applied in UMTS and broadband OFDM systems \cite{Bumman06}. As the corresponding peak-to-average power ratio (PAPR) of wide band signals is relatively high, the input power level of Doherty PAs has to be backed off in order to remain within the linear operating region. To deal with the nonlinear distortion of the Doherty PA, the digital predistortion (DPD) technique \cite{Morgan06} is often applied, which compensates the potential distortion through the BB processing before the power amplification. Combining the DPD technique with the Doherty architecture, the overall PA efficiency can be improved by more than 50\% according to the latest literature \cite{Pednekar18}.

To reduce the coupling loss of feeder and cooling systems, an architectural breakthrough for BSs is to introduce the separated BBU-RRU configuration \cite{Wang09} as shown in Figure~\ref{fig:net_model}, where the RRU part is installed near radio antennas and connected back to the BBU part through optical fibers. As the traditional cable feeder from BS to tower-mounted antennas incurs more than 3dB attenuation of wireless signals, the separated architecture can shorten the distance between the RRU and the antennas, which greatly reduces the feeder loss, $\sigma_{feed}$. In addition, as the RRU is equipped in the outdoor environment, the extra cooling system for them is no longer needed, as a result, the cooling loss, $\sigma_{cool}$, can be reduced as well. A natural extension of the separated BBU-RRU architecture is to stack the neighboring BBUs together to form a BBU pool and share the information and signal processing resources as well as the cooling systems, which is often referred to as cloud RAN (C-RAN). Although the deployment cost of optical fibers between the BBUs and the RRUs may increase, the cost reduction from site rent and maintenance, cooling systems, and BBU resources sharing will be more significant. Moreover, by implementing the BBU function into the general purpose computing platform and employing a unified protocol in the BBU-RRU link, such as {\em common public radio interface} (CPRI) \cite{Pfeiffer15}, cooperative transmission among different RRUs can be easily configured via centralized processing and the power consumption can be further reduced via statistical multiplexing of the BBU and the RRU pools.

Green radio techniques for RAN from different domains can be combined together to achieve higher EE, for example merging C-RAN and antennas switch-off using group sparse beamforming \cite{Shi14}, albeit some of them may conflict with each other. Even if the detailed analysis regarding hybrid solutions cannot be presented here, we summarize the main characteristics of green radio RAN schemes in Table~\ref{tab:GR_RAN}.

\begin{table*} [ht]
\centering
\caption{Summary of Green Radios for Radio Access Networks.}
\label{tab:GR_RAN}
\footnotesize
\begin{tabular}{c c c c c }
\toprule
Green Radio Schemes & Domain & Related Parameters & Power Saving Gains & Standard Impacts \\
\midrule
RF Chain Switch-off & Spatial & $N_{\textrm{TX}}$ & High & No \\
\midrule
Hybrid Precoder & Spatial & $N_{\textrm{TX}}$ & Medium to High & No \\
\midrule
Cell Switch-off & Spatial & $N_{\textrm{TX}}$ & High & No \\
\midrule
Cell DTX \& MBSFN & Time & $P_{sleep}$ & Medium & No \\
\midrule
Extended Cell DTX & Time & $P_{sleep}$ & Medium to High & Yes (ABS) \\
\midrule
Bandwidth Reduction & Frequency & $P_{o}$ & Low & No \\
\midrule
Energy Efficient Scheduling & Frequency & $P_{o}$ & Low & No \\
\midrule
mmWave and Unlicensed bands & Frequency & $P_{o}$ & Low to Medium & Yes (mmWave, LTE-U) \\
\midrule
Advance PA Architecture & Engineering & $\eta_{PA}$ & High & No \\
\midrule
BBU RRU Separation & Engineering & $\sigma_{feed}, \sigma_{cool}$ & High & No \\
\midrule
C-RAN & Engineering & $P_{BB}, \sigma_{cool}$ & Medium & Yes (CPRI) \\
\bottomrule
\end{tabular}
\end{table*}

\section{Advanced Green Terminals} \label{sect:at}
With the vision to connect everything through Internet, the {\em Internet of Things} (IoT) communications are recognized as a key player in future wireless networks. In addition to the conventional passive communication or sensing capabilities of traditional terminals, the newly defined ``things'', including smart meters, radio-frequency identification (RFID) tags, or even autonomous vehicles, require more intelligence through efficient communications, computing, and storage. As the large scale information exchange is necessary for various IoT applications nowadays, green IoT \cite{Shaikh17} for advanced terminals becomes essential to reduce the carbon footprints and the greenhouse gas emissions.

\subsection{D2D Communications}
Most of IoT applications, including cooperative monitoring, sensing, and decision require local information sharing among different IoT devices. Since the conventional uplink-downlink transmission via BSs consumes extra power over the air, direct communications among adjacent devices, also known as D2D communication \cite{Yu11,Feng13,Feng14}, has been proposed as an energy efficient solution for IoT networks and adopted in 3GPP D2D proximity services \cite{Seo16}. With the additional D2D transmission mode of advanced terminals, the related EE problems have been investigated under the centralized coordination scenario and the distributed sharing scenario.

Mode switching for D2D communications in cellular networks can be realized through the control plane signaling. In order to maximize EE, a unified framework for EE evaluation with mode switching has been developed in \cite{Feng15}. From \cite{Feng15}, the coexistence of D2D transmission and regular cellular transmission with careful power management is more preferable for EE improvement. In \cite{Chen15}, a joint optimization of social and physical networks for cooperative D2D transmission has been studied. By exploiting the social trust and social reciprocity, the network EE can be further improved.

The centralized coordination often involves significant signaling overhead for system information sharing, such as channel state information, which motivates us to consider the distributed sharing scenario. For instance, a reverse iterative combinatorial auction game is formulated to handle the EE maximization problem for the downlink D2D communication scenario in \cite{Xu131} and a nontransferable coalition game is applied for the uplink case in \cite{Wu14}. Both of the above two schemes utilize the game theoretic formulation to reduce the information exchange requirements and show significant EE improvement.

\subsection{Vehicle-to-Vehicle Communications}

\begin{figure}
\centering
\includegraphics[width = 3.4 in]{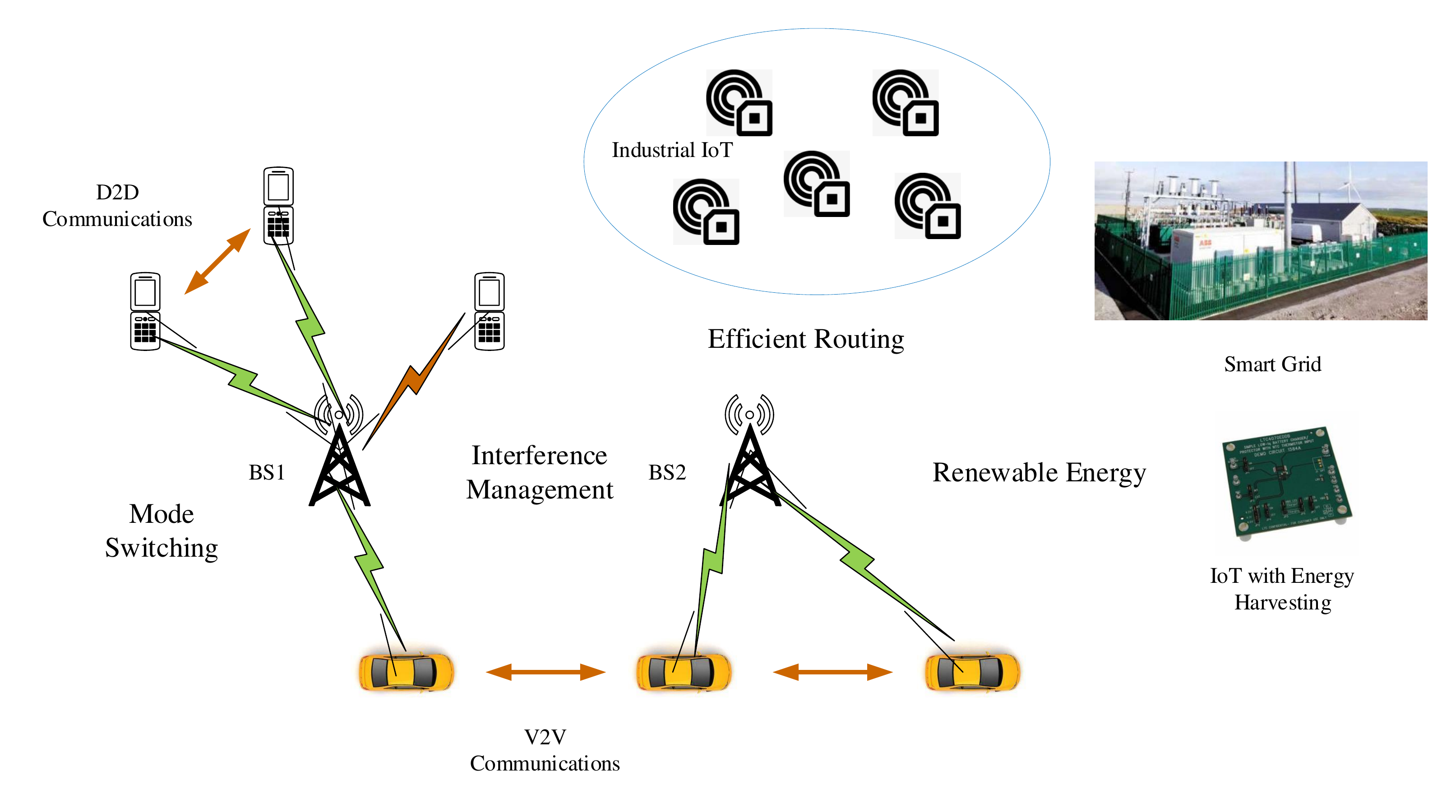}
\caption{An overview of green radios for IoT scenarios, including device-to-device, vehicle-to-vehicle, and other types of machine-to-machine communication systems.}
\label{fig:D2D_V2V}
\end{figure}

Although the green radio technologies developed in D2D communications can be extended to vehicle-to-vehicle (V2V) communications as shown in Figure~\ref{fig:D2D_V2V}, there are still several challenges to improve EE in this case.
\begin{itemize}
\item The high mobility of vehicles makes the D2D mode switching strategy difficult to implement in the V2V scenario.
\item To guarantee the low latency and high reliability requirement for V2V communications, other energy efficient schemes in both centralized and distributed coordination scenarios with higher computational complexity may not be applicable.
\item As the handover processes frequently happen in V2V communications, the redundant packet forwarding and retransmission cause additional performance loss in EE.
\end{itemize}

With the above understanding, the feasibility of using D2D communications for V2V transmission have been investigated in \cite{Cheng15}. From \cite{Cheng15}, distributed interference control and predictive resource allocation are necessary to improve EE in V2V communications.

Despite of the fact that high mobility of vehicles incurs design challenges in green radio areas, several unique characteristics in V2V communications can be explored. A typical example is that as vehicles can only operate on the roads, the prediction of terminal movement and potential V2V transmission is much easier, which facilitates more reliable energy efficient resource allocation for green cities \cite{Zhou181}.

\subsection{IoT Communications with Renewable Energy}

\begin{figure}
\centering
\includegraphics[width = 3.4 in]{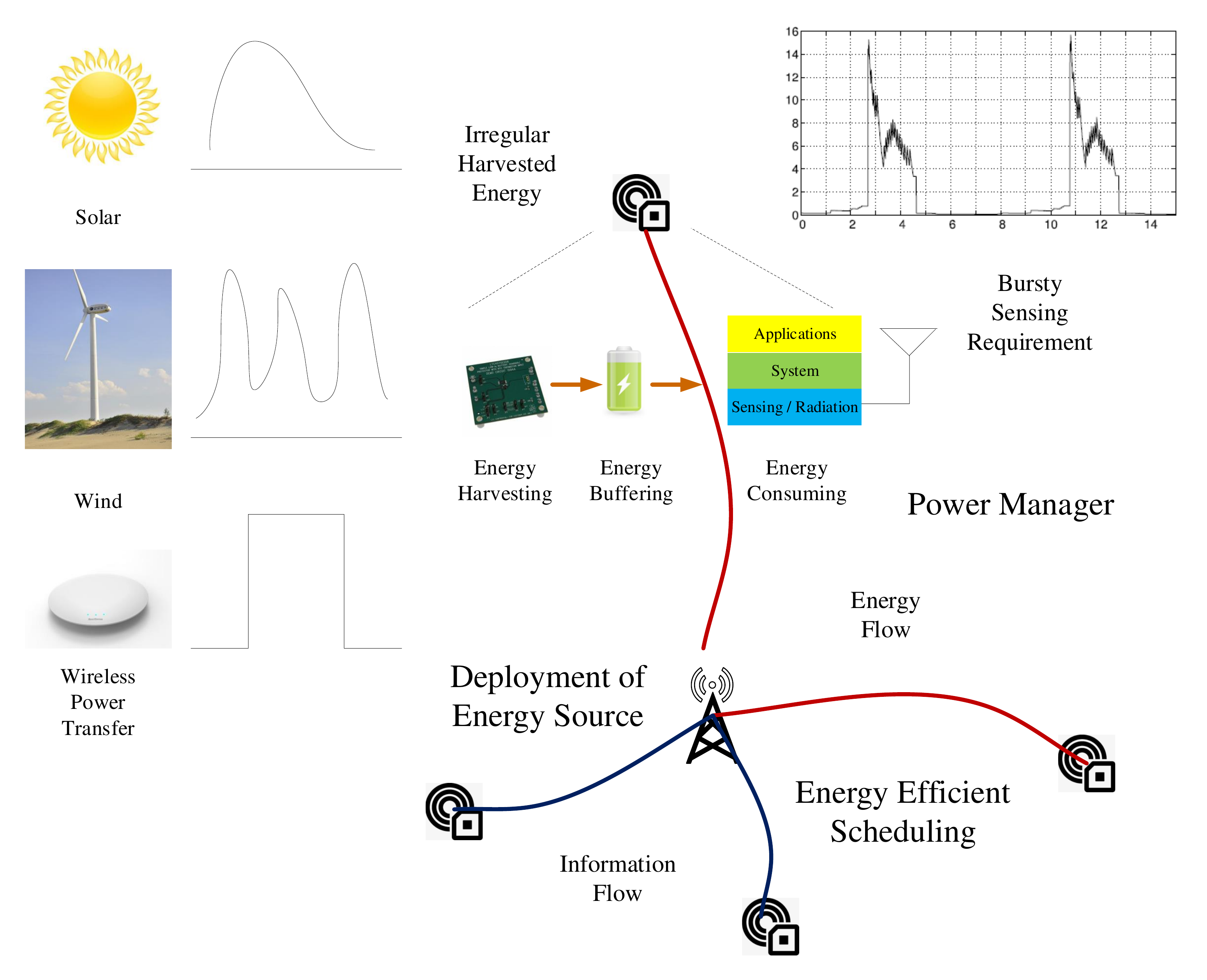}
\caption{An illustration of green IoT with renewable energy.}
\label{fig:renew_en}
\end{figure}

Another important breakthrough for green IoT is to incorporate the renewable energy, which combines energy harvesting, buffering, and consuming as an integrated form as shown in Figure~\ref{fig:renew_en}. The corresponding power models, as shown in \cite{Martinez15}, include a new component called power manager to balance the harvested energy and the consumed energy. Since the renewability is often unpredictable at the terminal side, an energy efficient protocol to communicate with the renewable energy source will be desirable \cite{Kamalinejad15}. For example, a low power wake-up radio is equipped to monitor the environment and activates regular transceiver chains when renewable energy sources are detected. A more attractive approach to combine the energy harvesting behavior with social networking characteristics has been proposed in \cite{Jiang16}, where the interactions among the social domain, the device domain, and the energy domain are emphasized and utilized for EE improvement.

To make the renewable energy source more reliable and predictable, a dedicated source for energy harvesting, also known as wireless power transfer, has been proposed in \cite{Sample11} and commonly adopted in the state-of-the-art commercial smartphones, such as iPhone X. Although the dedicated energy source needs to modify the infrastructure, it can cooperatively communicate with the energy harvesting terminals and deliver the corresponding energies whenever necessary. Meanwhile, if the BS is equipped with multiple antennas, the wireless power transfer process is able to combine with the wireless information transmission, which brings significant EE improvement as shown in \cite{Zhang13}.

As summarized in \cite{Ejaz17}, in addition to energy harvesting protocol and receiver design, the deployment of dedicated energy sources, the multi-path energy routing, and the efficient scheduling mechanisms to guarantee coverage and service quality are crucial for the EE improvement for IoT communications with energy harvesting.

\subsection{Other Types of IoT Communications}
Other types of energy efficient IoT communications usually identify the unique challenges in some specific application areas. The smart grid aims to bring intelligence to the conventional energy distribution and control system, which is often regarded as a paradigm shift for efficient energy utilization. To deal with the potential energy issues in remote monitoring, controlling, and optimization for the smart grid, IoT technology decomposes the associated tasks into several decentralized control procedures via energy efficient peer-to-peer communications \cite{Bui12}. Industrial IoT (IIoT) is another example, where the design challenge is the organization and interactions of the densely deployed sensor nodes. To address this issue, a three layer nodes control framework has been proposed in \cite{Wang16} to balance the sensing requirement and the controllable sleep mode scheduling, which can reduce the overall energy consumption by half. To expand the coverage of IoT communication for large scale sensing tasks, several narrow band technologies, such as narrow band IoT (NB-IoT) \cite{Wang17} and low power wide area (LPWA) \cite{Xiong15}, have been proposed to maintain the total energy consumption. In Table~\ref{tab:adv_term}, we summarize the green radio schemes for advanced green terminals.

\begin{table*} [ht]
\centering
\caption{Summary of Green Radios for Advanced Green Terminals.}
\label{tab:adv_term}
\footnotesize
\begin{tabular}{c c c}
\toprule
Type of Communications & Green Radio Schemes & Reference \\
\midrule
D2D & mode switching between D2D transmission and regular cellular transmission & \cite{Feng15} \\
\midrule
D2D & centralized coordination information exchange reduction & \cite{Xu131,Wu14} \\
\midrule
V2V & distributed interference control and predictive resource allocation & \cite{Cheng15} \\
\midrule
V2V & energy efficient resource allocation with the infrastructure (road) constraint & \cite{Zhou181} \\
\midrule
IoT w. Renewable Energy & energy harvesting schemes, efficient deployment, routing and scheduling & \cite{Liu07,Lin09} \\
\midrule
Others & energy efficient peer-to-peer communication, NB-IoT, and LPWA & \cite{Bui12,Wang16,Wang17,Xiong15}\\
\bottomrule
\end{tabular}
\end{table*}

The above energy efficient solutions for advanced terminals, by no means complete, provide some guidance for designing green IoT technologies. However, there are still several important issues to be addressed in the near future.
\begin{itemize}
\item The definition of EE for IoT networks needs to be reinvestigated, especially when the harvested energy and the control signaling of massive IoT devices in the standby/sleep modes are taken into consideration. Meanwhile, the processing needed for running energy harvesting protocols may itself consume too much energy and the processors needed may be more expensive than what IoT promises.
\item The interference prediction and management for D2D communication links are challenging as well and the distributed interference coordination is desired.
\item The joint optimization for various IoT applications by adapting heterogeneous communication protocols, including cellular, WiFi, or even ZigBee/bluetooth, is another promising direction in the green IoT design.
\item Energy efficient schemes for mission critical IoT applications, such as ultra reliable low latency transmission for V2V communications, are yet to be investigated based on the existing literature.
\end{itemize}

\section{Future Green Networks} \label{sect:fn}
The primary design goal of future network architectures is to offer ubiquitous and sufficient amount of wireless data services with guaranteed quality. To fulfill the aforementioned divergent requirements of future wireless communications and support new types of IoT devices, future networks should incorporate heterogenous network configurations and future radio access technology evolution, where the software defined network (SDN) as well as the network function virtualization (NFV) are the key enablers. Meanwhile, to reduce the round trip delay in URLLC and mMTC scenarios, mobile edge computing (MEC) is often utilized to offload mission critical applications and provide efficient localized services. With the above techniques for future network architectures, energy efficient design is necessary for sustainable network operation, especially when massive number of antennas or cells are deployed.

\subsection{Network Virtualization}

The network virtualization technologies, including the aforementioned SDN and NFV, are built on top of some common infrastructure (such as servers). The entire power consumption of virtualized networks is usually higher than the traditional RAN architecture, as additional control and configuration processes are required to accommodate the flexibility. However, the network virtualization facilitates several important features that may allow more aggressive design for green radios. For example, the separation of control and data planes is often regarded as one of the most important features for SDN, which can be applied to design functionality separated green mobile networks \cite{Xu132}. A similar concept can be extended to energy plane to deal with the renewable energy entities as well, where an software defined approach to manage the energy routing network and the hierarchical energy control architecture can provide significant EE improvement from \cite{Zhong16}. Another important feature for the network virtualization is the locally centralized control, through which the energy efficient traffic prediction and management is possible. By dynamically programming the network functions and routing protocols according to the varying traffic profiles, substantial energy saving gain can be realized \cite{Sezer13}.

Proper placement of the network virtualization entities, including SDN controllers, SDN switches, and other end devices is another potential degree of freedom for green radio design although it still suffers from the scalability issue as well as the interconnection issue with legacy networks. A mathematical model to incorporate the traffic requirements and the location of virtualization entities has been developed in \cite{Sallahi15} while the energy efficient placement is still open.

\subsection{MEC}

\begin{figure}
\centering
\includegraphics[width = 3.4 in]{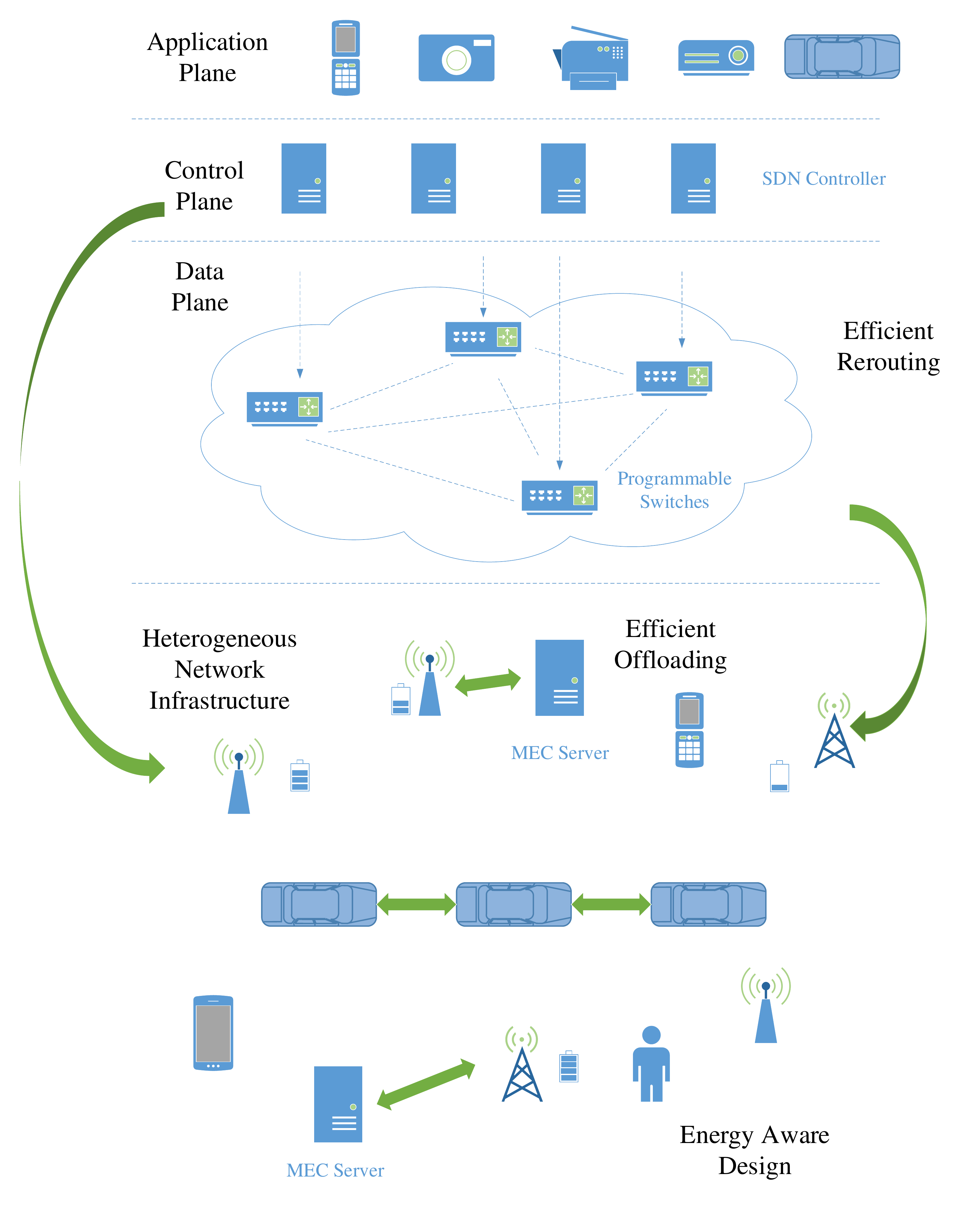}
\caption{An overview of energy efficient architectures for future networks.}
\label{fig:FN}
\end{figure}

A promising future network architecture to partially address the scalability issue of massive wireless devices is to utilize the MEC framework, where the data requests from mobile terminals are offload to local MEC servers rather than core networks. With the aid of MEC, the information processing and exchange are mainly performed at the edge nodes. Hence, the corresponding transmission energy can be saved as the frequent intercommunication with core networks is unnecessary. To minimize the energy consumption of MEC on top of that, efficient traffic offloading and rerouting schemes are usually regarded as valuable solutions. For instance, the multiple access nature of wireless communications has been exploited in \cite{Zhang16}, where the energy minimized computing task offloading schemes have been developed to improve EE in 5G heterogeneous networks. The extension to jointly utilize communication and computation budgets for energy efficient design has been proposed in \cite{You17}, where a threshold-based offloading strategy is proved to be optimal in terms of weighted energy minimization. Furthermore, depending on different applications, such as augmented reality \cite{Ali17} or vehicular services \cite{Shojafar18}, the above schemes are tailored to improve the EE performance under specific QoS constraints at the edge nodes.

Despite of the energy efficient task offloading and resource scheduling, it is also worthwhile to note that green radios can be achieved via proper deployment of MEC. As the tradeoff relation between the number of mobile edge nodes and the improved EE performance is still unknown, the energy efficient MEC deployment policy design still suffers from limited scientific guidance, which we believe will attract significant research attentions in the near future.

\subsection{Joint BB and RF Design}

\begin{figure}
\centering
\includegraphics[width = 3.4 in]{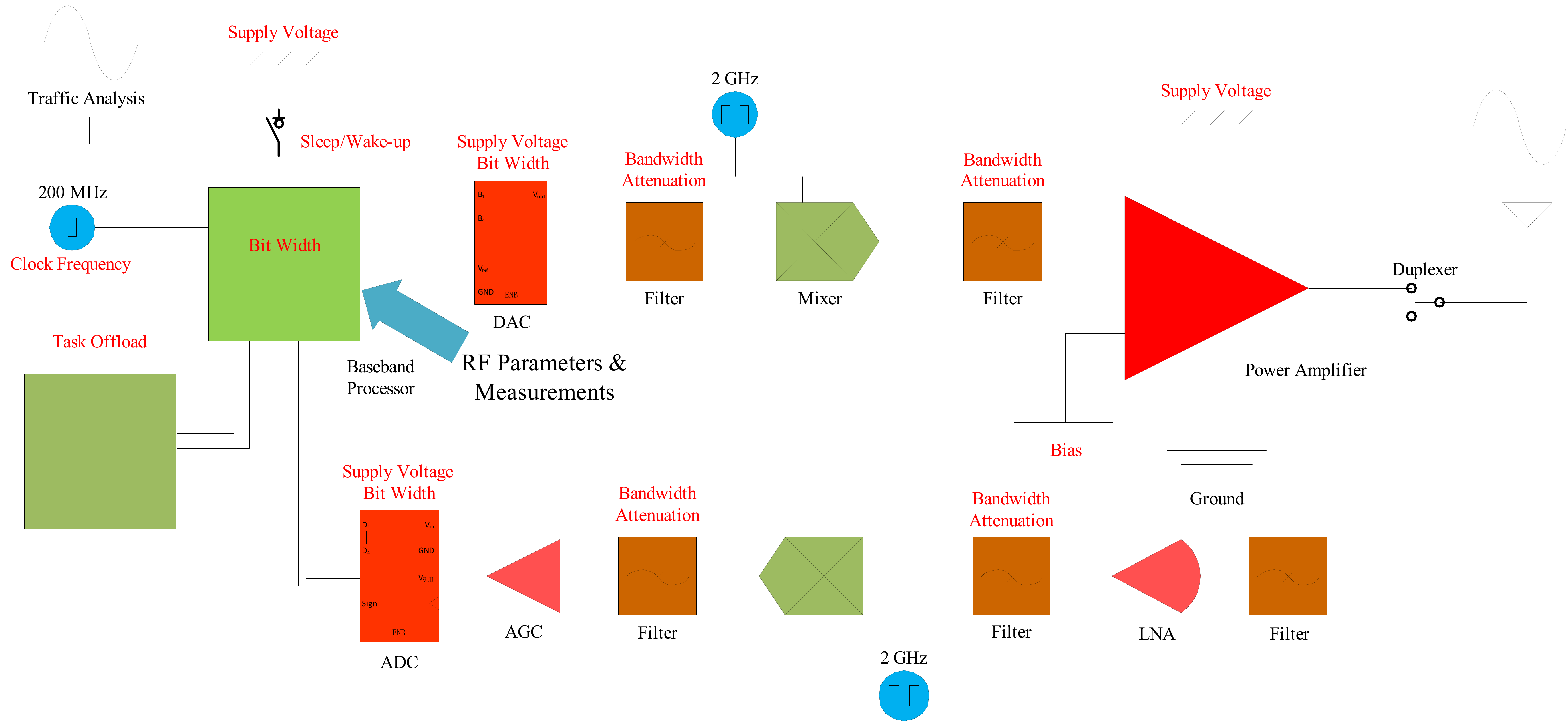}
\caption{An overview of BB and RF structures for future mobile networks.}
\label{fig:BB-RF}
\end{figure}

With more complicated network architecture and additional reconfigurability requirements to satisfy software defined property, the implementation of future networks faces greater challenges on the power consumption. As a novel approach to deal with this issue, the wireless designers start to consider the characteristics of RF components in the original system design and improve the implementation efficiency via joint BB and RF processing \cite{Zhang19}. A successful practice in massive MIMO design is to reduce the resolution level of analog-to-digital converters (ADCs), for uplink detection as shown in \cite{Choi16}, where one bit quantization for real and imaginary parts of received signals provides sufficient detection accuracy with only 1-3 dB SNR loss. In \cite{Liang16}, the concept of low-resolution quantization has been extended to mixed ADC structure and the resolutions of ADCs are dynamically configured according to the BB processing requirement. In addition to the resolutions of ADCs, various parameters of RF components, such as the supply voltage and the bias voltage, can be adjusted to reduce the power consumption based on the traffic analysis in the BB processing. In Figure~\ref{fig:BB-RF}, we summarize the tunable parameters in the current RF chains, which can be jointly optimized with more intelligent control in the BB implementation.

\subsection{Cognitive Radio and Network Intelligence}
The last promising area to impact future energy efficient communications is to explore the cognitive dimension as summarized in \cite{Gur11}. As the cognitive technique includes a complete learning cycle of sensing, planning, deciding, and acting, it can be fully utilized to improve the system EE from different layers. In the NET and the above layers, the cognitive concept has been used to facilitate interactions with real-time varying traffic and energy sources, such as dynamic traffic offloading \cite{Chen151} and smart grid powered wireless networks \cite{BU12}. Meanwhile, this type of operations will further allow some high-power core network entities to go to sleep when the network traffic can be maintained through some low-power devices \cite{Gur11}. In the MAC and PHY layers, the cognitive functionality has been proposed to incorporate with the traditional resource allocation schemes in both heterogeneous networks \cite{Xie12} and cooperative transmission \cite{Huang15} environments.

In addition to the conventional cognitive radio schemes, the machine learning concept has also been applied to improve the network intelligence recently. As summarized in \cite{Jiang17}, machine learning can be widely adopted the next generation wireless network design. For example, the supervised learning with regression models can be applied to estimate or predict the handover behaviors to dynamically switch-off cells for energy saving \cite{Shen16}, and the unsupervised learning, such as K-means clustering, is a powerful tool to utilize the network cooperation for energy efficient heterogenous networks \cite{Chen17}. To provide more accurate decision making for energy efficient applications, reinforcement learning has been widely used for energy harvesting related problems, such as in \cite{Zheng15}. On top of that, a transfer actor-critic learning framework has been propose in \cite{Li14}, where the transferable learning expertise in historical periods or neighboring regions can be utilized to improve the current network energy saving strategy.

Nevertheless, green radios for future networks require more intelligence from all the possible layers \cite{Zhou18}. Although the above energy efficient schemes cover several layers in the wireless communication networks, the end-to-end framework \cite{Qin19} with sufficient network intelligence is still open. With more powerful tools, such as machine learning, we believe future networks shall be capable of supporting advanced green features through self-learning, self-adapting, and self-evolving.

\section{Conclusion} \label{sect:conc}

In this article, we have summarized the achievements in green radios in its first 20 years, including terminal and RAN sides. Rather than only emphasizing the theoretical results of energy efficient communications, green radio technologies in the engineering aspects with significant EE saving gain are discussed as well. Based on the investigation of green radios in the first 20 years, we believe future green radio technologies will continue to evolve in the next decades, including energy efficient schemes with new scenarios (e.g., URLLC and mMTC), new optimization framework (e.g., machine learning based optimization tools), and new hardware dynamics (e.g., BB and RF characteristics). Nevertheless, green radio schemes in the first 20 years have provided a fundamental basis for designing future energy efficient wireless networks.

\section*{Acknowledgement}
This work was in part supported by the National Natural Science Foundation of China (NSFC) Grants under No. 61701293 and No. 61871262, the National Science and Technology Major Project Grants under No. 2018ZX03001009, the Huawei Innovation Research Program (HIRP), and research funds from Shanghai Institute for Advanced Communication and Data Science (SICS).

\bibliographystyle{IEEEtran}
\bibliography{IEEEabrv,20year_green}
\begin{IEEEbiography}[{\includegraphics[width=1in,height=1.25in,clip,keepaspectratio]{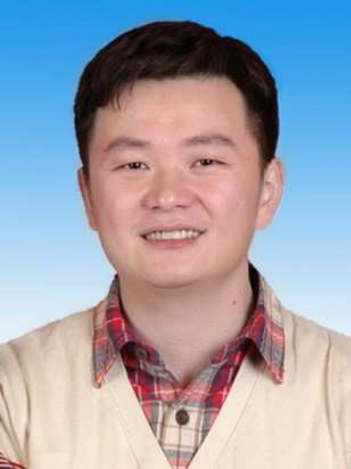}}]{Shunqing Zhang}
(S'05-M'09-SM'14) received the B.S. degree from the Department of Microelectronics, Fudan University, Shanghai, China, in 2005, and the Ph.D. degree from the Department of Electrical and Computer Engineering, Hong Kong University of Science and Technology, Hong Kong, in 2009.

He was with the Communication Technologies Laboratory, Huawei Technologies, as a Research Engineer and then a Senior Research Engineer from 2009 to 2014, and a Senior Research Scientist of Intel Collaborative Research Institute on Mobile Networking and Computing, Intel Labs from 2015 to 2017. Since 2017, he has been with the School of Communication and Information Engineering, Shanghai University, Shanghai, China, as a Full Professor. His current research interests include energy efficient 5G/5G+ communication networks, hybrid computing platform, and joint radio frequency and baseband design. He has published over 60 peer-reviewed journal and conference papers, as well as over 50 granted patents. He has received the National Young 1000-Talents Program and won the paper award for Advances in Communications from IEEE Communications Society in 2017.
\end{IEEEbiography}

\begin{IEEEbiography}[{\includegraphics[width=1in,height=1.25in,clip,keepaspectratio]{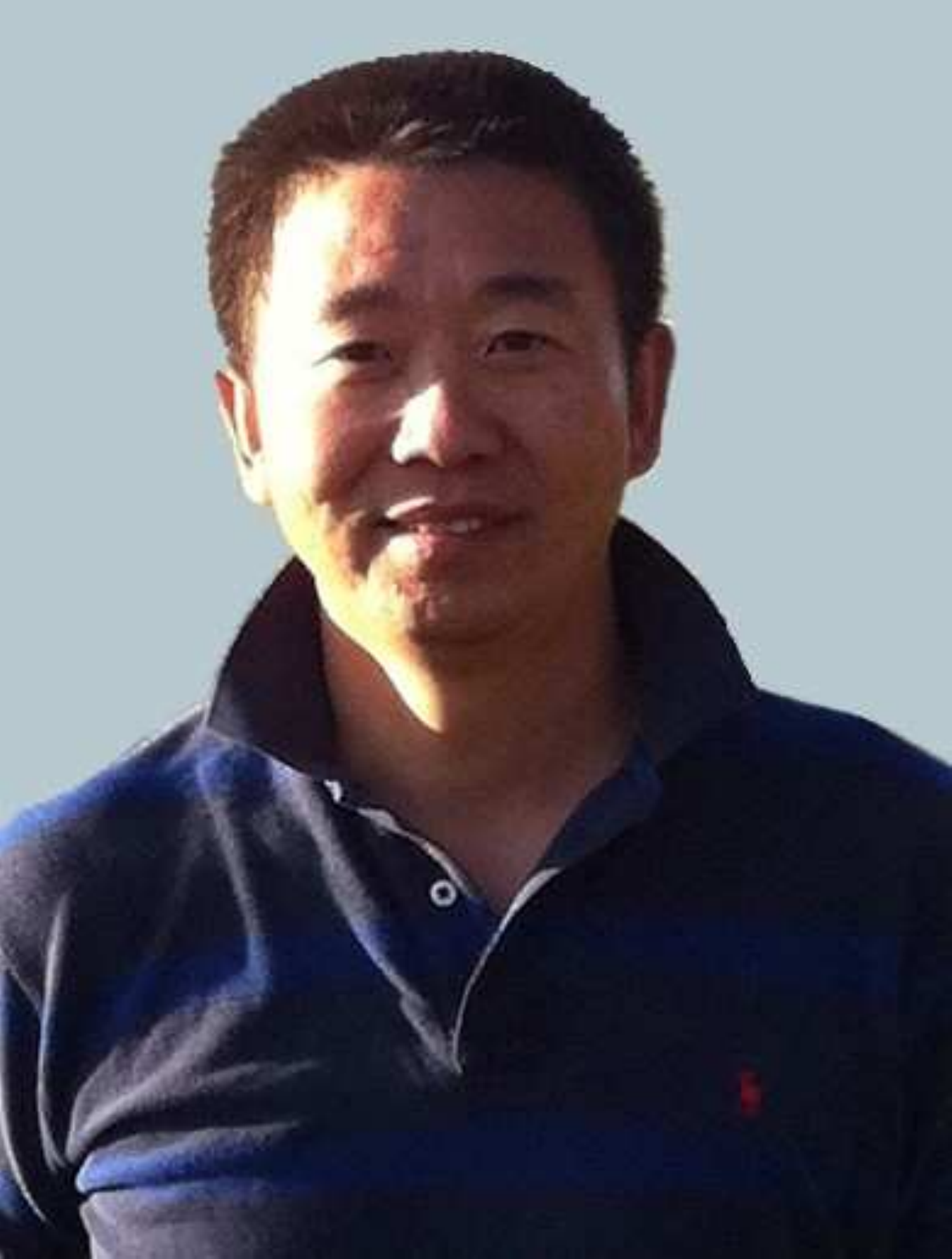}}]{Shugong Xu}
(M'98-SM'06-F'16) graduated from Wuhan University, China, in 1990, and received his Master degree in Pattern Recognition and Intelligent Control from Huazhong University of Science and Technology (HUST), China, in 1993, and Ph.D. degree in EE from HUST in 1996.He is professor at Shanghai University, head of the Shanghai Institute for Advanced Communication and Data Science (SICS). He was the center Director and Intel Principal Investigator of the Intel Collaborative Research Institute for Mobile Networking and Computing (ICRI-MNC), prior to December 2016 when he joined Shanghai University. Before joining Intel in September 2013, he was a research director and principal scientist at the Communication Technologies Laboratory, Huawei Technologies. Among his responsibilities at Huawei, he founded and directed Huawei's green radio research program, Green Radio Excellence in Architecture and Technologies (GREAT). He was also the Chief Scientist and PI for the China National 863 project on End-to-End Energy Efficient Networks. Shugong was one of the co-founders of the Green Touch consortium together with Bell Labs etc, and he served as the Co-Chair of the Technical Committee for three terms in this international consortium. Prior to joining Huawei in 2008, he was with Sharp Laboratories of America as a senior research scientist. Before that, he conducted research as research fellow in City College of New York, Michigan State University and Tsinghua University. Dr. Xu published over 100 peer-reviewed research papers in top international conferences and journals. One of his most referenced papers has over 1400 Google Scholar citations, in which the findings were among the major triggers for the research and standardization of the IEEE 802.11S. He has over 20 U.S. patents granted. Some of these technologies have been adopted in international standards including the IEEE 802.11, 3GPP LTE, and DLNA. He was awarded ``National Innovation Leadership Talent'' by China government in 2013, was elevated to IEEE Fellow in 2015 for contributions to the improvement of wireless networks efficiency. Shugong is also the winner of the 2017 Award for Advances in Communication from IEEE Communications Society. His current research interests include wireless communication systems and Machine Learning.
\end{IEEEbiography}
\begin{IEEEbiography}[{\includegraphics[width=1in,height=1.25in,clip,keepaspectratio]{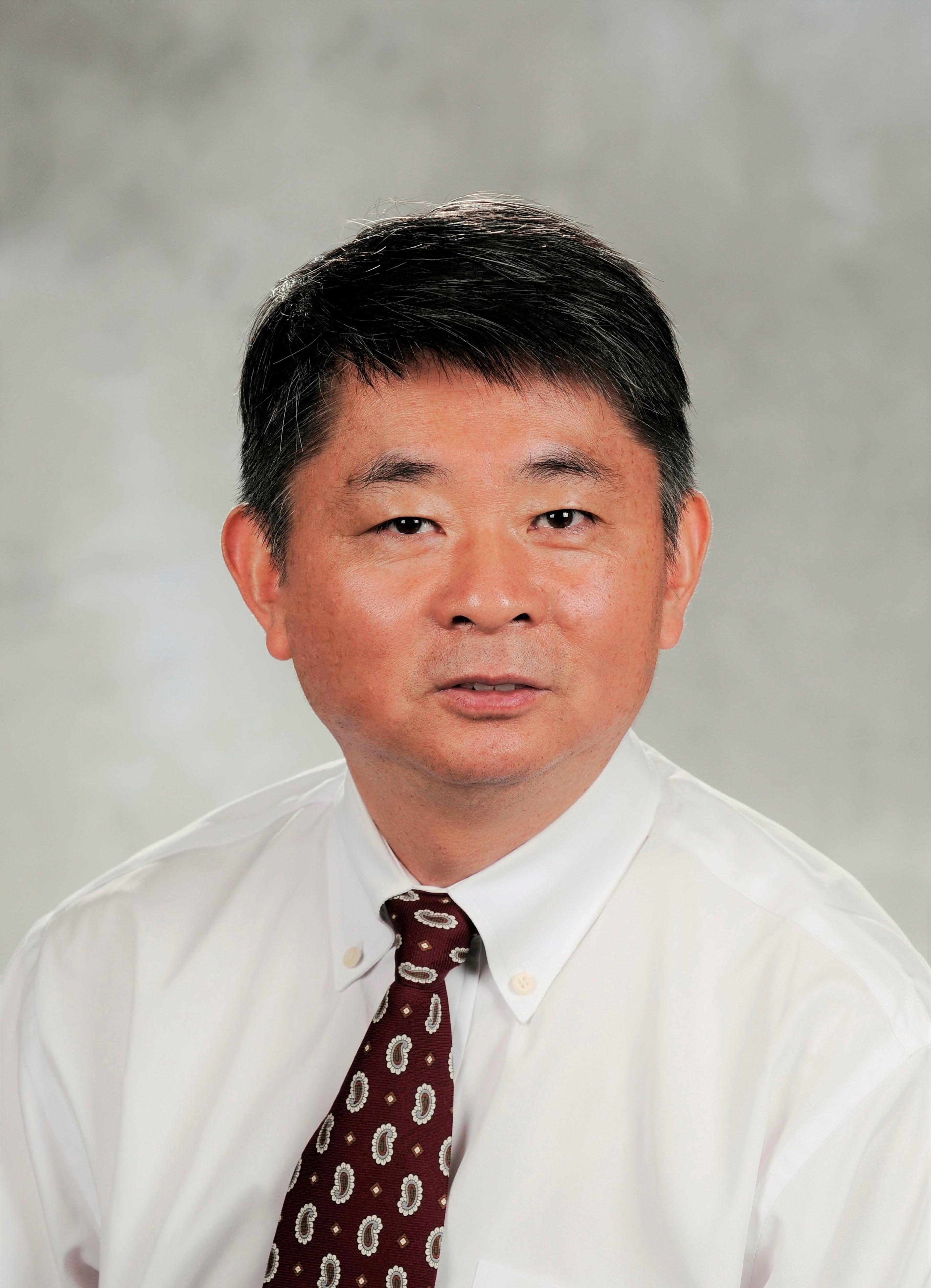}}]{Geoffrey Ye Li}
(S'93-M'95-SM'97-F'06) received his B.S.E. and M.S.E. degrees in 1983 and 1986, respectively, from the Department of Wireless Engineering, Nanjing Institute of Technology, Nanjing, China, and his Ph.D. degree in 1994 from the Department of Electrical Engineering, Auburn University, Alabama.

He was a Teaching Assistant and then a Lecturer with Southeast University, Nanjing, China, from 1986 to 1991, a Research and Teaching Assistant with Auburn University, Alabama, from 1991 to 1994, and a Post-Doctoral Research Associate with the University of Maryland at College Park, Maryland, from 1994 to 1996. He was with AT\&T Labs - Research at Red Bank, New Jersey, as a Senior and then a Principal Technical Staff Member from 1996 to 2000. Since 2000, he has been with the School of Electrical and Computer Engineering at Georgia Institute of Technology as an Associate Professor and then a Full Professor.

His general research interests include statistical signal processing and machine learning for wireless communications. In these areas, he has published over 500 journal and conference papers in addition to over 40 granted patents. His publications have been cited around 37,000 times and he has been recognized as the World¡¯s Most Influential Scientific Mind, also known as a Highly-Cited Researcher, by Thomson Reuters almost every year. He was awarded IEEE Fellow for his contributions to signal processing for wireless communications in 2005. He won 2010 IEEE ComSoc Stephen O. Rice Prize Paper Award, 2013 IEEE VTS James Evans Avant Garde Award, 2014 IEEE VTS Jack Neubauer Memorial Award, 2017 IEEE ComSoc Award for Advances in Communication, and 2017 IEEE SPS Donald G. Fink Overview Paper Award. He also received 2015 Distinguished Faculty Achievement Award from the School of Electrical and Computer Engineering, Georgia Tech. He has been involved in editorial activities for over 20 technical journals for the IEEE, including founding Editor-in-Chief of IEEE 5G Tech Focus. He has organized and chaired many international conferences, including technical program vice-chair of IEEE ICC¡¯03, technical program co-chair of IEEE SPAWC¡¯11, general chair of IEEE GlobalSIP¡¯14, technical program co-chair of IEEE VTC¡¯16 (Spring) and general co-chair of IEEE VTC¡¯19 (Fall) .

\end{IEEEbiography}
\begin{IEEEbiography}[{\includegraphics[width=1in,height=1.25in,clip,keepaspectratio]{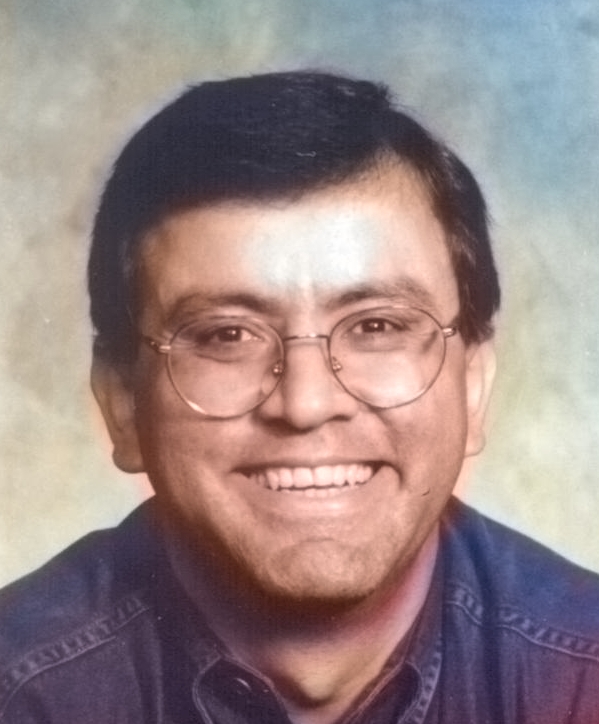}}]{Ender Ayanoglu}
(S'82-M'85-SM'90-F'98) received the Ph.D. degree in electrical engineering from Stanford University, Stanford, CA, USA, in 1986. He was with the Communications Systems Research Laboratory, Holmdel, NJ, USA, part of AT\&T Bell Laboratories, until 1996, and Bell Laboratories, Lucent Technologies, from 1996 to 1999. From 1999 to 2002, he was a Systems Architect with Cisco Systems, Inc., San Jose, CA. Since 2002, he has been a Professor with the Department of Electrical Engineering and Computer Science, University of California, Irvine, CA, where he served as the Director of the Center for Pervasive Communications and Computing and the Conexant-Broadcom Endowed Chair from 2002 to 2010. His past accomplishments include invention of 56K modems, characterization of wavelength conversion gain in wavelength-division multiplexed systems, and diversity coding. From 2000 to 2001, he served as the Founding Chair of the IEEE-ISTO Broadband Wireless Internet Forum, an industry standards organization. He served on the Executive Committee of the IEEE Communications Society¡¯s Communication Theory Committee from 1990 to 2002 and the Chair from 1999 to 2002. From 1993 to 2014, he was an Editor of the IEEE TRANSACTIONS ON COMMUNICATIONS. He served as the Editor-inChief of the IEEE TRANSACTIONS ON COMMUNICATIONS from 2004 to 2008 and the IEEE JOURNAL ON SELECTED AREAS IN COMMUNICATIONS series on Green Communications and Networking from 2014 to 2016. Since 2014, he has been a Senior Editor of the IEEE TRANSACTIONS ON COMMUNICATIONS. Since 2016, he has been serving as the Founding Editor-in-Chief of the IEEE TRANSACTIONS ON GREEN COMMUNICATIONS AND NETWORKING. He received the IEEE Communications Society Stephen O. Rice Prize Paper Award in 1995, the IEEE Communications Society Best Tutorial Paper Award in 1997, and the IEEE Communications Society Communication Theory Technical Committee Outstanding Service Award in 2014.
\end{IEEEbiography}
\end{document}